\documentclass[manuscript]{aastex5}

\usepackage{lscape}
\usepackage{graphicx}
\usepackage{here}

\begin{document}


\title{Discovery of a Hot Corino in the Bok Globule B335}


\author{Muneaki~Imai\altaffilmark{1}, Nami~Sakai\altaffilmark{2}, Yoko~Oya\altaffilmark{1}, Ana~L$\acute{\mathrm{o}}$pez-Sepulcre\altaffilmark{1,3}, Yoshimasa~Watanabe\altaffilmark{1}, Cecilia~Ceccarelli\altaffilmark{4,5}, Bertrand~Lefloch\altaffilmark{4,5}, Emmanuel~Caux\altaffilmark{6,7}, Charlotte~Vastel\altaffilmark{6,7}, Claudine~Kahane\altaffilmark{4,5}, Takeshi Sakai\altaffilmark{8}, Tomoya~Hirota\altaffilmark{9}, Yuri~Aikawa\altaffilmark{10}}
\and 
\author{Satoshi~Yamamoto\altaffilmark{1,11}}

\altaffiltext{1}{Department of Physics, The University of Tokyo, 7-3-1, Hongo, Bunkyo-ku, Tokyo 113-0033, Japan}
\altaffiltext{2}{The Institute of Physical and Chemical Research (RIKEN), 2-1, Hirosawa, Wako-shi, Saitama 351-0198, Japan}
\altaffiltext{3}{Institut de Radioastronomie Millim$\acute{\mathrm{e}}$trique (IRAM), 300 rue de la Piscine, 38406 Saint-Martin-d'H$\grave{\mathrm{e}}$res, France}
\altaffiltext{4}{Universite de Grenoble Alpes, IPAG, F-38000 Grenoble, France}
\altaffiltext{5}{CNRS, IPAG, F-38000 Grenoble, France}
\altaffiltext{6}{Universite de Toulouse, UPS-OMP, IRAP, Toulouse, France}
\altaffiltext{7}{CNRS, IRAP, 9 Av. Colonel Roche, BP 44346, F-31028 Toulouse Cedex 4, France}
\altaffiltext{8}{Department of Communication Engineering and Informatics, Graduate School of Informatics and Engineering, The University of Electro-Communications, Chofugaoka, Chofu, Tokyo 182-8585, Japan}
\altaffiltext{9}{National Astronomical Observatory of Japan, Osawa, Mitaka, Tokyo 181-8588, Japan}
\altaffiltext{10}{Center for Computational Science, University of Tsukuba, Tsukuba, Ibaraki 305-8577, Japan}
\altaffiltext{11}{Research Center for the Early Universe, The University of Tokyo, 7-3-1, Hongo, Bunkyo-ku, Tokyo 113-0033, Japan}

\begin{abstract}
We report the first evidence of a hot corino in a Bok globule. This is based on the ALMA observations in the 1.2 mm band toward the low-mass Class 0 protostar IRAS 19347+0727 in B335. Saturated complex organic molecules (COMs), CH$_3$CHO, HCOOCH$_3$, and NH$_2$CHO, are detected in a compact region within a few 10 au around the protostar. Additionally, CH$_3$OCH$_3$, C$_2$H$_5$OH, C$_2$H$_5$CN, and CH$_3$ COCH$_3$ are tentatively detected. Carbon-chain related molecules, CCH and c-C$_3$H$_2$, are also found in this source, whose distributions are extended over a few 100 au scale. On the other hand, sulfur-bearing molecules CS, SO, and SO$_2$, have both compact and extended components. Fractional abundances of the COMs relative to H$_2$ are found to be comparable to those in known hot-corino sources. Though the COMs lines are as broad as 5-8 km s$^{-1}$, they do not show obvious rotation motion in the present observation. Thus, the COMs mainly exist in a structure whose distribution is much smaller than the synthesized beam (0\farcs58 $\times$ 0\farcs52).
\end{abstract}



\section{Introduction}
During the last decade, it has been established that chemical compositions of protostellar cores show significant diversity even among those in similar evolutionary stages (Class 0/I). One distinct case is hot corino chemistry, which is characterized by the rich existence of saturated complex-organic molecules (COMs: molecules consisting of six or more atoms) such as HCOOCH$_3$ and CH$_3$OCH$_3$. A prototypical case is IRAS 16293-2422 in Ophiuchus ({\it e.g.} \citealt{cazaux2003}; \citealt{bottinelli2004}; \citealt{kuan2004}). Another distinct case is warm-carbon-chain chemistry (WCCC), which is characterized by abundant unsaturated organic molecules (carbon-chain molecules and their related species) in a warm and dense region around a protostar (\citealt{sakai2008}; \citeyear{sakai2009}; \citeyear{sakai2010}). A prototypical source is IRAS 04368+2557 in L1527 in the Taurus molecular cloud. These two cases have exclusive chemical compositions: carbon-chain molecules are deficient in hot corinos, while COMs are deficient in the WCCC sources. Since the chemical composition can reflect the past evolutionary history of the source, a duration time of a starless core phase after shielding of the interstellar UV radiation is proposed as a possible origin of the chemical diversity \citep{sakaiandyamamoto2013}. A longer duration time of the starless core phase tends to result in hot corino chemistry, while a shorter duration time leads to WCCC. Environmental effects such as location in a larger-scale molecular cloud,  influences of nearby protostellar sources, and strength of UV radiation field ({\it e.g.} \citealt{watanabe2012}; \citealt{lindberg2015}; \citealt{spezzano2016}) may also affect the chemical composition. 

Since the chemical composition of the protostellar cores would reflect not only current physical conditions but also evolution histories, exploration of a total picture of the chemical diversity is a fundamental issue both for astrochemistry and star-formation studies. However, only a few hot corinos and only a few WCCC sources have so far been detected definitively, and hence, it is still essential to study chemical compositions of other protostars to unveil the total picture. In particular, it is important to study the chemical composition of a protostellar core in an isolated condition, which is thought to be free from various environmental effects caused by other protostars.

B335 is an ideal target for this purpose. It is a representative Bok globule \citep{keene1980}, which harbors the Class 0 low-mass protostar IRAS 19347+0727. Its distance and bolometric luminosity are reported to be 100 pc \citep{olofsson2009} and 0.72 $L_{\sun}$ \citep{evans2015}, respectively.  This source is regarded as the best  "test-bed" for detailed studies of simple models of star-formation ({\it e.g.} \citealt{hirano1988}; \citealt{hirano1992}; \citealt{zhou1993}; \citealt{chandler1993}; \citealt{wilner2000}; \citealt{harvey2001}). \citet{yen2015} recently conducted the ALMA observation at a resolution of 0\farcs34 $\times$ 0\farcs28, and reported that the protostar in B335 has no Keplerian disk with a radius of 10 au or larger. \citet{evans2015} also reported on the basis of their ALMA observations that HCN and HCO$^+$ lines show absorption features against continuum, which are well reproduced by the model of inside-out collapse. In addition to these physical studies, its chemical composition at a few 1000 au scale has been observed and modeled \citep{evans2005}, where fundamental molecules such as CO, CN, HCO$^+$, HCN, HNC, N$_2$H$^+$, and H$_2$CO were studied. For full understandings of chemical evolution to the protoplanetary disk, the chemical composition in the closest vicinity of the protostar has to be explored. With this in mind, we here report the first chemical characterization of this source at a few 10 au scale with ALMA.

\section{Observation}
We conducted the 1.2 mm (250 GHz) observation (Band 6) of B335 with ALMA (Cycle 2) on May 18, 2015. In total, 37 antennas were used in the observation. The field center is $\alpha$(J2000)=19$^\mathrm{h}$37$^{\mathrm{m}}$0\fs93, $\delta$(J2000)=7\degr34\arcmin9\farcs9. Minimum and maximum baselines are 15 k$\lambda$ and 444 k$\lambda$, respectively. A primary beam size is 23\farcs6, and a synthesized beam size is 0\farcs58 $\times$ 0\farcs52 (P.A. = 75\degr). We used 16 spectral windows whose bandwidth and channel spacing are 58.6 MHz and  61.027 kHz, respectively. The velocity resolution is 0.1404 km s$^{-1}$. On-source integration time was ~34 minutes, which resulted in the rms noise of 3-7 mJy beam$^{-1}$. Titan was observed as a flux calibrator. J1955+1358 and J1751+0939 were observed as a phase calibrator and a bandpass calibrator respectively. Self-calibration was not applied for simplicity. The calibration accuracy is 10 \%.

We used CASA for the data reduction. A continuum image was prepared by averaging line-free channels. Maps of the spectral line emission were obtained by CLEANing the dirty images (Briggs robust = 0.5) after subtracting the continuum directly from the visibility data. 

\section{Results}

\subsection{Dust Continuum}
Figure~1a shows the 1.2 mm continuum map. The peak position is determined by the Gaussian fit as: ($\alpha_{2000}$, $\delta_{2000}$) =  (19$^{\mathrm{h}}$37$^{\mathrm{m}}$0\fs90, 7\degr34\arcmin9\farcs62). The continuum emission consists of compact and extended components. A deconvolved size of the compact component of the continuum emission is 0\farcs43 $\times$  0\farcs28 with the position angle of 23$\pm$15\degr. The extended component has a structure extended along the north-south direction which is perpendicular to the outflow direction \citep{hirano1988}. The total integrated flux is 92.8$\pm$2.1 mJy. This result is consistent with the total continuum flux (compact + extended) of 87.5$\pm$2.8 mJy at 1.3 mm reported by \citet{yen2015}.

\subsection{Detected Molecules and Their Distribution}
Figure~2 shows the 16 spectral windows observed toward the continuum peak position. In this observation, B335 is found to be rich in molecular lines. We assigned 31 spectral lines using the Cologne Database for Molecular Spectroscopy (CDMS) \citep{muller2005} and Jet Propulsion Laboratory (JPL) \citep{pickett1998} databases assuming the systemic velocity of 8.34 km s$^{-1}$ \citep{yen2015}, while 5 lines are left unidentified. They are summarized in Table~1, as well as the line parameters obtained by the Gaussian fit. The most noteworthy result is the detection of CH$_3$CHO, HCOOCH$_3$, and NH$_2$CHO. Additionally, CH$_3$OCH$_3$, C$_2$H$_5$OH, C$_2$H$_5$CN, and CH$_3$COCH$_3$ are tentatively detected, each of which is identified only by a single faint feature. These saturated COMs are characteristic of hot corinos  and hot cores of star-forming regions. This is the first detection of COMs in this source. In addition to saturated COMs, the carbon-chain molecule CCH and the carbon-chain related molecule c-C$_3$H$_2$ are also detected.  

Spectral line profiles are different from molecule to molecule, as shown in Figure~2 and Table~1. Based on the FWHM width of the line ($\Delta v$), they are roughly classified into the following three categories: \\
(1) Broad lines ($\Delta v \gtrsim$ 5 km s$^{-1}$): Twenty-one lines of the 11 molecular species including tentatively detected ones are classified in this group. They are mostly saturated COMs and related molecules (Table~1). They do not show absorption below the continuum emission level ({\it i.e.}, the baseline)   \\
(2) Narrow lines ($\Delta v < 1.6$ km s$^{-1}$): CCH and c-C$_3$H$_2$ show the narrow line width with a  double-peak structure due to absorption by the foreground gas. Five lines of these two species are classified into this category. \\
(3) Intermediate lines ($\Delta v \simeq$ 3.0 km s$^{-1}$): Sulfur-bearing molecules such as SO, $^{34}$SO, SO$_2$, $^{34}$SO$_2$ and CS belong to this category. These species have an intermediate line width between the first and second categories. Five lines of the above five species are classified into this category.  

Figures~1(b-i) show moment 0 maps of HCOOCH$_3$, NH$_2$CHO, HNCO, CH$_3$CHO, HCOOH, CCH, CS, and SO. Molecular distributions are different among the above three categories. The moment 0 maps of the COM related lines (HCOOCH$_3$, NH$_2$CHO, HNCO, CH$_3$CHO, and HCOOH), which have broad line widths, reveal a compact distribution concentrated around the protostar (Figures~1(b-f)). They are not resolved with the synthesized beam of this observation (0\farcs58 $\times$ 0\farcs52, P.A. = 55\degr): the deconvolved sizes of the emitting region (FWHM) for HCOOCH$_3$ and NH$_2$CHO are (0\farcs42 $\pm$ 0\farcs07) $\times$ (0\farcs31 $\pm$ 0\farcs08) and (0\farcs42 $\pm$ 0\farcs06) $\times$ (0\farcs30 $\pm$ 0\farcs07), respectively. According to the previous studies \citep{hirano1988}, the outflow is extended along the east-west direction, and the protostellar disk/envelope system would have a nearly edge-on configuration with respect to the line-of-sight ($i \sim10 \degr$). Nevertheless, the rotation motion cannot be identified in the COMs spectrum. This result is consistent with the results reported by \citet{yen2015} and  \citet{evans2015}. Hence, the broad line width of the COMs means that the COMs mainly exist in  a structure whose distribution much smaller than the synthesized beam.

On the other hand, the CCH emission, which shows the narrow line width, is extended over a few 100 au scale from the protostar (Figure~1g). Such an extended distribution is consistent with the self-absorption feature seen in the spectra. A part of the extended component would trace an outflow cavity wall, whose direction (east-west) is consistent with the previous studies \citep{hirano1988}. Although this source harbors a hot corino, carbon-chain related species can be observed in the protostellar core. The moment 0 maps of the sulfur-bearing molecules, showing the intermediate line width, reveal the compact distribution (Figures~1(h,i)), although a weak extended component can also be seen. 

\subsection{Derivation of Column Density and Fractional Abundance}
We derive the beam-averaged column densities of HCOOCH$_3$, CH$_3$OCH$_3$, CH$_3$CHO, NH$_2$CHO, HNCO, c-C$_3$H$_2$, SO$_2$, HC$_3$N, HCOOH, CH$_3$COCH$_3$, SO, and CS toward the continuum peak, assuming the local thermodynamic equilibrium (LTE) and optically thin conditions:
\begin{equation}
N(\mathrm{X})=U(T) \frac{3 k_\mathrm{B} W}{8 \pi^3 \nu S \mu^2} \exp \left(\frac{E_\mathrm{u}}{k_\mathrm{B} T}\right),
\end{equation}
where $U(T)$ denotes the partition function of the molecule at the temperature $T$, $W$ the integrated intensity, and $E_\mathrm{u}$ the upper state energy. Since a single line or multiple lines with similar upper state energies are observed in this study, the column densities are calculated for the excitation temperature of 100 K. This temperature is a typical excitation temperature of COMs in the hot corino source IRAS 16293-2422 (\citealt{richard2013}, \citealt{jaber2014}, \citealt{oya2016}), and is also close to the mass-weighted dust temperature \citep{kauffmann2008}  at a roughly 0\farcs5 beam (111 K) reported for B335 by \citet{evans2015}. Although the detection of CH$_3$OCH$_3$ and CH$_3$COCH$_3$ is tentative in this study, we calculate their column densities, for comparison with other sources. The results are shown in Table~2.

To derive the fractional abundances relative to H$_2$, the beam-averaged column density of H$_2$ are estimated by using the following equation:
\begin{equation}
N(\mathrm{H}_2) = \frac{F(\nu) N_\mathrm{A}}{\nu ^3 \kappa_{\nu} \Omega \rho (2h/c^2)} [ \exp(h \nu / k_\mathrm{B} T_\mathrm{d})-1],
\end{equation}
where $\nu$ denotes the frequency, $F(\nu)$ the peak integrated flux of dust continuum emission, $\kappa_{\nu}$ the mass absorption coefficient with respect to the gas mass,  $T_\mathrm{d}$ the dust temperature, $\Omega$ the solid angle of the synthesized beam, $\rho$ the average molecular weight (2.33) in the atomic mass unit, and $N_\mathrm{A}$ the Avogadro's number \citep{ward2000}. The mass absorption coefficient at 1.2 mm is calculated to be 0.0068 cm$^2$ g$^{-1}$ by using the average value of $\kappa_{\nu}$ at 850 $\mu$m ($1.48 \times 10^{-2}$) with the $\beta$ index of 2.38, which are reported for B335 by \citet{shirley2011}. Then, the column density of H$_2$ is derived to be $(5.66 \pm 0.11) \times 10^{23}\ \mathrm{cm}^{-2}$ for the dust temperature of 100 K.  Using the H$_2$ column density, the fractional abundances of the observed molecules relative to H$_2$ are evaluated, as summarized in Table~2. Here, we simply assume that the dust temperature equals to the LTE temperature. If the line-of-sight depth of the molecular distribution ($L$) were the same as the FWHM of the continuum peak ($L \simeq 1.1 \times 10^{15}$ cm), the H$_2$ density is roughly estimated to be $n(\mathrm{H}_2) \sim 5\times10^8$ cm$^{-3}$. Such a relatively high density justifies the LTE assumption employed in derivation of the column densities.

In the following section, we discuss the abundances of the saturated COMs and their related species. The abundances of sulfur-bearing molecules and carbon-chain molecules will be discussed in a separate publication. (Sakai et al. in preparation)

\section{Discussion}
In this observation, we detected the lines of CH$_3$CHO, HCOOCH$_3$, and NH$_2$CHO, and tentatively detected the lines of CH$_3$OCH$_3$, C$_2$H$_5$OH, C$_2$H$_5$CN, and CH$_3$COCH$_3$. These COM lines show a broad line width and a very compact distribution around the protostar. The fractional abundance of HCOOCH$_3$ relative to H$_2$ is evaluated to be $4.6 \times 10^{-9}$, which is comparable to that reported for the prototypical hot corino IRAS 16293-2422 ($9 \times 10^{-9}$) \citep{jaber2014} and NGC 1333 IRAS 4A ($1.4 \times 10^{-9}$) \citep{taquet2015}. The fractional abundance of CH$_3$OCH$_3$ in B335 ($3.4 \times 10^{-9}$) is smaller than that reported for IRAS 16293-2422 ($4 \times 10^{-8}$) \citep{jaber2014}, but is slightly higher than that for NGC1333 IRAS 4A. Thus, B335 is confirmed to be rich in COMs. The HCOOCH$_3$ abundance is also comparable to that in the outflow shocked region L1157 B1 \citep{sugimura2011}. On the other hand, the fractional abundances of HCOOCH$_3$ and CH$_3$OCH$_3$ in B335 are much higher than those found in the prestellar core L1689B (10$^{-10}-10^{-9}$) \citep{bacmann2012}. This comparison indicates that the abundances of COMs are enhanced in the compact region near the protostar of B335, and hence, we conclude that B335 harbors a hot corino.

Among the various COMs, NH$_2$CHO is proposed to be a key species in pre-biological evolution \citep{saladino2012}. This molecule has been detected in hot corinos and hot cores in star forming regions (\citealt{bisschop2007}; \citealt{adande2013}; \citealt{kahane2013}). It is reported that the abundance of NH$_2$CHO shows a good correlation with that of HNCO, implying that these two species are related to each other in their production mechanisms \citep{ana2015}. In B335, the NH$_2$CHO/HNCO ratio is 0.024, which is almost comparable to the range of the ratios found in star forming regions (0.03-0.25). Hence, the positive correlation between the abundances of these two species indeed holds in B335.

Detection of acetone (CH$_3$COCH$_3$) is tentative in this source. If the 248.682 GHz line originates from acetone, as in the case of NGC1333 IRAS 4A (L$\acute{\mathrm{o}}$pez-Sepulcre et al. in preparation), the fractional abundance of acetone in B335 is determined to be $0.8 \times 10^{-9}$. The acetone abundance is lower than the value reported toward the acetone peaks in Orion KL (hot core) ($(0.4-1.6) \times 10^{-8}$ by \citealt{friedel2005}). According to the interferometric observations toward Orion KL by \citet{friedelandsnyder2008} and \citet{peng2013}, acetone shows different distribution from the other O-bearing COMs (HCOOCH$_3$ and CH$_3$OCH$_3$), and its distribution tends to be similar to the N-bearing COMs (C$_2$H$_5$CN). In B335, the distribution of acetone is similar to those of the other COMs, and we cannot find any specific trend in its distribution unlike the Orion KL case, probably because of the insufficient angular resolution in this study. Definitive identification of acetone in this source with multiple lines and its high resolution imaging are awaited.

So far, hot corino sources have been found in large cloud complexes with active star formation: IRAS 16293-2422 is in the Ophiuchus molecular cloud complex \citep{cazaux2003}, NGC1333 IRAS 4A, IRAS 4B, and IRAS 2A \citep{bottinelli2004, sakai2006, jorgensen2005} in the Perseus molecular cloud complex, Serpens SMM4 in the Serpens molecular cloud \citep{oberg2011}, and HH212 in the Orion molecular cloud \citep{codella2016}. In contrast, B335 is the first hot corino source identified in a Bok globule isolated from a large molecular cloud complex. It is generally thought that a star forming region in a molecular cloud complex would be affected by various activities of the other protostars in the same cloud complex. Likewise, the chemical composition of the protostellar core would also be affected by such environmental effects. In contrast, B335 is thought to be almost free from them. Its chemical composition could be regarded as a 'standard' template for chemical compositions of isolated protostellar cores. It will also be very useful for comparison with chemical model calculations in order to understand chemical evolution during star formation.

Finally, we note an implication to the protostellar mass in this source. Although the compact distributions of the COMs are not spatially resolved, we may be able to constrain the size of the emitting region and the protostellar mass on the basis of the following rough argument.   Among the lines which have a compact distribution and a broad line width, the brightest one is the HNCO line, excluding tentative detections and blended lines. Its peak flux and the line width are 72.8 mJy beam$^{-1}$ and 5.19 km s$^{-1}$ respectively. The peak flux corresponds to the brightness temperature of 4.2 K for the 0\farcs58 $\times$ 0\farcs52 (58 au $\times$ 52 au) beam.  Judging from the distribution and the line width, HNCO is most likely present in the same region where the COMs exist. If this emission comes from such a compact region, the actual brightness temperature corrected for the beam dilution should be higher. If COMs and related species including HNCO are assumed to be distributed within the region whose average temperature is 100 K (a typical hot corino temperature), the brightness temperature has to be lower than 100 K. From this condition, the diameter of the emitting region is estimated to be 11 au or larger. To explain the line width at this range of the radius, the protostellar mass is estimated to be 0.04 $M_\odot$ or higher, if the motion responsible to the line width is Keplerian. It is 0.02 $M_\odot$ or higher, if the motion is infall. These mass estimates are consistent with that reported by \citet{yen2015} ($> 0.05\ M_\odot$).

\acknowledgments
The authors are grateful for a reviewer of this paper for invaluable comments. This paper makes use of the following ALMA data: ADS/JAO.ALMA\#2013.1.01102.S. ALMA is a partnership of ESO (representing its member states), NSF (USA) and NINS (Japan), together with NRC (Canada), NSC and ASIAA (Taiwan), and KASI (Republic of Korea), in cooperation with the Republic of Chile. The Joint ALMA Observatory is  operated by ESO, AUI/NRAO and NAOJ. This study is financially supported by KAKENHI (21224002, 25400223, and 25108005). The authors acknowledge the financial support by JSPS and MAEE under the Japan-France integrated action programme (SAKURA: 25765VC).




\begin{figure*}
	\figurenum{1}
	\includegraphics[bb = 0 0 900 600, width = 18 cm]{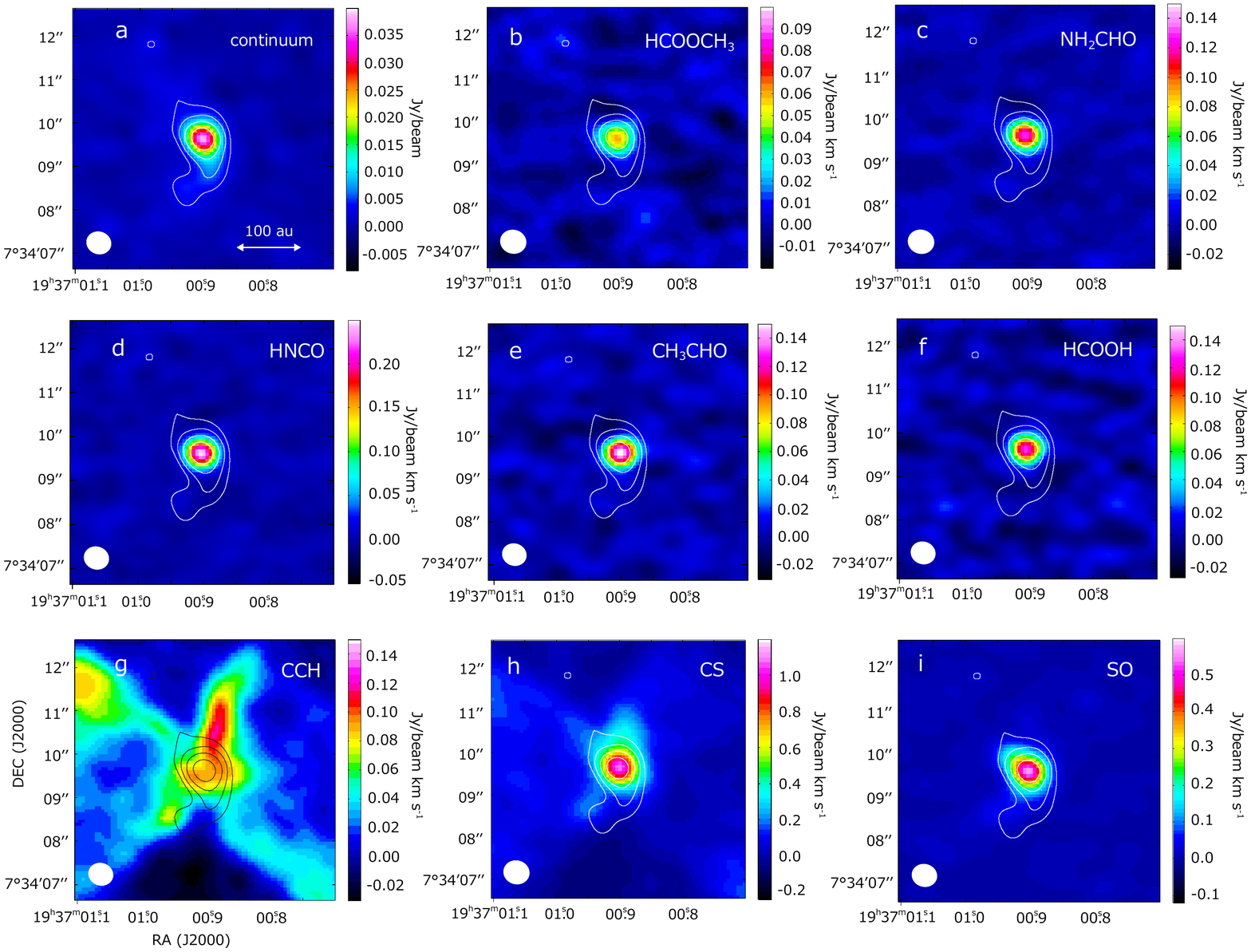}
	\caption{The continuum map and the moment 0 maps of HCOOCH$_3$ ($20_{5,16}-19_{5,15}$), NH$_2$CHO ($12_{0,12}-11_{0,11}$), HNCO ($12_{0,12}-11_{0,11}$), CH$_3$CHO ($14_{1,14}-13_{1,13}$ E), HCOOH ($12_{0,12}-11_{0,11}$), CCH ($N=3-2, J =5/2-3/2, F=3-2$), CS ($5-4$), and SO ($N_J =6_7-5_6$). The contours represent the continuum flux of 10, 20, 40, 80$\sigma$ levels, where $\sigma$ is 0.3 mJy beam$^{-1}$. Compared with the synthesized beamsize shown in the bottom left in each figure, distributions of COMs are not resolved. The velocity range is $5.06-11.67$ km s$^{-1}$ for HCOOCH$_3$, $5.20-10.22$ km s$^{-1}$ for NH$_2$CHO, $5.89-10.82$ km s$^{-1}$ for HNCO, $5.03-11.62$ km s$^{-1}$ for CH$_3$CHO, $2.85-13.66$ km s$^{-1}$ for HCOOH, $7.02-9.46$ km s$^{-1}$ for CCH, $6.28-10.54$ km s$^{-1}$ for CS, and $6.69-9.27$ km s$^{-1}$ for SO.}
\end{figure*}

\begin{landscape}
\begin{figure*}
\begin{center}
	\figurenum{2}
	\includegraphics[bb = 0 0 1000 550, width = 28 cm]{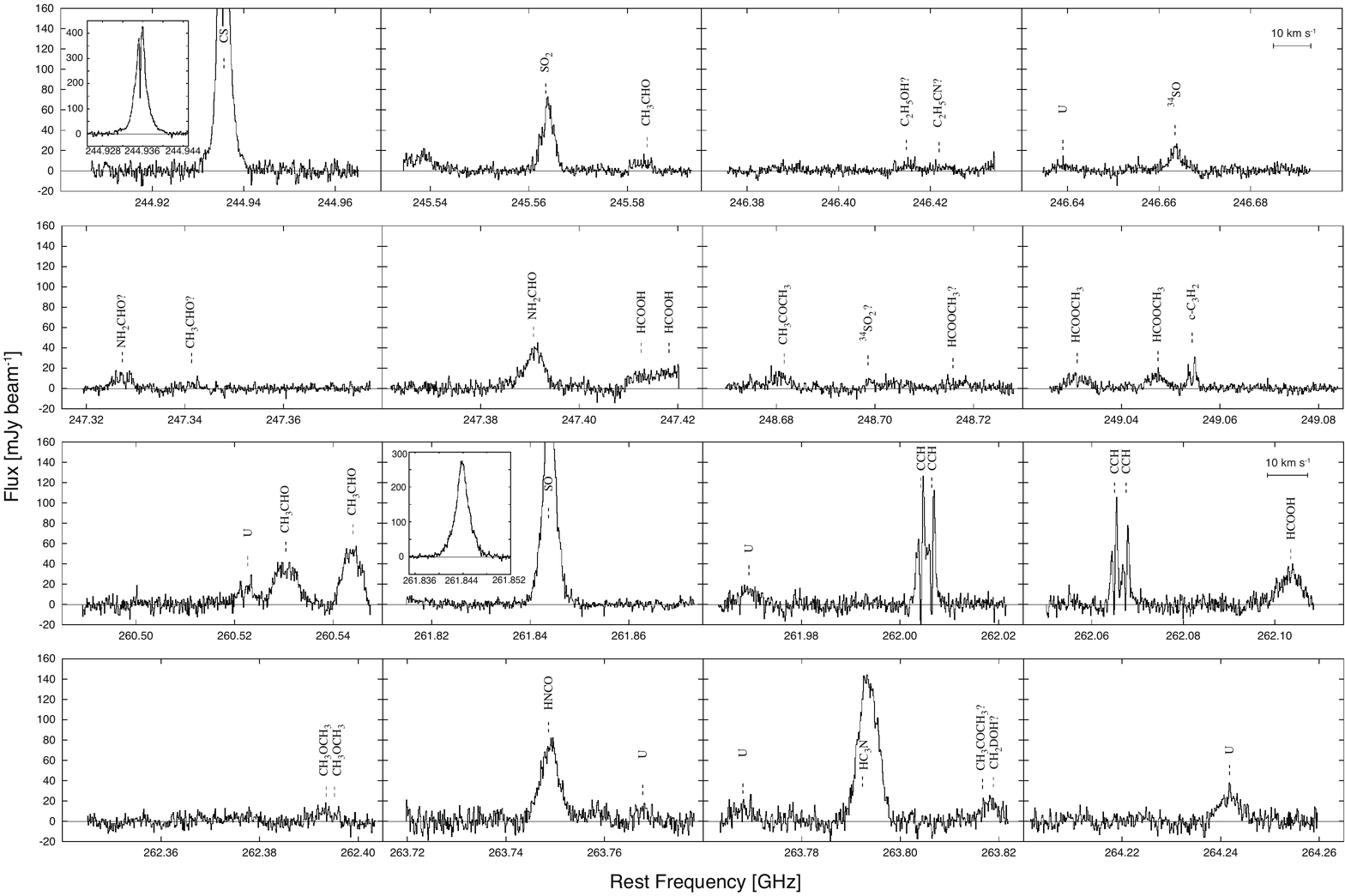}
	\caption{Spectrum toward the continuum peak of B335 observed with ALMA.}
\end{center}
\end{figure*}
\end{landscape}

\begin{deluxetable}{rrccccccccc}
\tabletypesize{\scriptsize}
\rotate
\tablecaption{Detected molecular lines}
\tablecolumns{11}
\tablenum{1}
\tablewidth{0pt}
\tablehead{
\colhead{Molecule\tablenotemark{a}} & \colhead{Transition} & \colhead{Frequency\tablenotemark{b}} & \colhead{$E_\mathrm{u}$} &
\colhead{$S\mu ^2$} & \colhead{$F_{\mathrm{integ}}$} & \colhead{$I$}	& \colhead{$\Delta v$} & \colhead{$v_\mathrm{lsr}$} & \colhead{rms} & \colhead{type\tablenotemark{c}}\\
\colhead{} & \colhead{} & \colhead{GHz} & \colhead{cm$^{-1}$} &  & 
\colhead{mJy beam$^{-1}$ km s$^{-1}$} & \colhead{mJy beam$^{-1}$} & \colhead{km s$^{-1}$} & \colhead{km s$^{-1}$} & \colhead{mJy beam$^{-1}$ ch$^{-1}$}
 }
\startdata
CS\tablenotemark{d}  & 5$-$4 & 244.935644 & 24.51 & 19.1 & 1301(4) & & & & 5.39 & I \\
SO$_2$ & $10_{3,7}-$10$_{2,8}$ & 245.563423 & 50.54 & 14.5 & 228(2) & 61.1(14) & 3.58(9) & 7.77(4) & 3.13 & I \\
CH$_3$CHO\tablenotemark{e} & $13_{0,13}-12_{0,12}$ E $v_\mathrm{t}=1$ & 245.583105 &  199.84 & 163 & 39(2) & 8.2(7) & 4.6(5) & 8.59(19) & 3.13 & B\\
C$_2$H$_5$OH? & $14_{3,11}-13_{3,10}$ & 246.414762 & 108.23 & 21.3 & 30(2) & 5.6(6) & 5.3(7) & 7.8(3) &3.17 & B\\
C$_2$H$_5$CN? & 28$_{2,27}-$27$_{2,26}$ & 246.421918 & 123.20 & 412 & 20(3) & 3.5(4) & 8(2) & 7.1(8) & 3.17 & B \\
U-line &  & 246.63899(19) & & & 20(3) & 5.5(6) & 4.5(5) & & 3.74 & \\
$^{34}$SO & 6$_5-$5$_4$ & 246.663470 & 34.68 & 11.4 & 92(3) & 20.1(9) & 3.9(2) & 8.17(8) & 3.74 & I \\
NH$_2$CHO?  & $12_{0,12}-11_{0,11}$ $v_{12}=1$ & 247.327322 & 343.26 & 156 & 56(2) & 10.8(8) & 5.3(5) & 8.82(18) & 3.24 & B \\
CH$_3$CHO?\tablenotemark{f} & 14$_{0,14}-$13$_{-1,13}$ E & 247.341332 & 66.56 & 11.3 & 15.1(19) & & & &3.24 & B \\
NH$_2$CHO & 12$_{0,12}-$11$_{0,11}$ & 247.390719 & 107.25 & 0.319 & 206(4) & 35.9(7) & 5.80(14) & 8.12(6) & 4.57 & B \\
HCOOH\tablenotemark{g} & 11$_{8}-$10$_{8}$ & 247.412546 & 190.76 & 20.9 & & & & & 4.57 & B \\
HCOOH\tablenotemark{g} & 11$_{7}-$10$_{7}$ & 247.418230 & 157.69 & 26.4 & & & & & 4.57 & B \\ 
CH$_3$COCH$_3$\tablenotemark{h} & $24_{2,23}-23_{1,22}$ EE & 248.681568 & 108.64 & 3050 & 52(3) & 9.7(6) & 7.5(5) & 10.2(2) &  3.78 & B\\
$^{34}$SO$_2$? & 13$_{1,13}-$12$_{0,12}$ & 248.698604 & 56.86  & 26.0 & 16(2) & 7.4(8) & 2.4(3) & 7.69 (13) & 3.78 & I \\
HCOOCH$_3$? & $20_{3,14}-19_{3,16}$ A $v_\mathrm{t}=1$& 248.715840 & 223.01 & 102 & 25(3) & 4.8(5) & 7.6(10) & 6.6(4) & 2.97 & B\\
HCOOCH$_3$ & 20$_{5,16}-$19$_{5,15}$ E & 249.031002 & 98.39 & 49.8 & 60(2) & 10.0(7) & 5.7(5) & 8.31(19) & 2.97 & B \\
HCOOCH$_3$ & 20$_{5,16}-$19$_{5,15}$ A & 249.047428 & 98.39 & 49.3 & 56(2) & 11.9(6) & 5.0(3) & 8.61(13) & 2.97 & B \\
c-C$_3$H$_2$\tablenotemark{d} & 5$_{2,3}-$4$_{3,2}$ & 249.054368  & 24.51 & 19.1 & 41.1(14) &  & & & 2.97 & N \\
U-line &  & 260.5227(3) &  &  & 67(3) & 15.5(12) & 5.2(9) & & 5.45 & \\
CH$_3$CHO & 14$_{1,14}-$13$_{1,13}$ E  & 260.530403 & 67.00 & 176 & 223(4) & 37.0(9) & 6.16(19) & 8.37(7) & 5.45 & B \\
CH$_3$CHO & 14$_{1,14}-$13$_{1,13}$ A & 260.544020 & 66.95 & 176 & 265(4) & 54.2(11) & 4.80(12) & 8.59(5) & 5.45 & B\\
SO & $N_J = 6_7-$5$_6$ & 261.843684 & 33.05 & 16.4 & 847(3) & 245(3) & 3.13(4) & 8.09(2) & 3.30 & I\\
U-line & & 261.96932(16) & & & 69(5) & 13.5(11) & 4.4(4) & & 5.65 &  \\
CCH\tablenotemark{d} & $N=3-2, J =7/2-5/2, F=4-3$ & 262.004260 & 17.48 & 194 & 136(3) & & & & 5.65 & N \\
CCH\tablenotemark{d} & $N=3-2, J =7/2-5/2, F=3-2$ & 262.006482 & 17.48 & 51.4 & 129(2) &  & & & 5.65 &  N \\
CCH\tablenotemark{d} & $N=3-2, J =5/2-3/2, F=3-2$ & 262.064986 & 17.49 & 1.63 & 100(3) & & & & 5.70 & N \\
CCH\tablenotemark{d} & $N=3-2, J =5/2-3/2, F=2-1$ & 262.067469 & 17.49 & 1.07 & 89(3) & & & &  5.70 & N \\
HCOOH & 12$_{0,12}-$11$_{0,11}$ & 262.103480 & 57.53 & 24.2 & 190(5) & 29.5(8) & 6.4(2) & 8.39(9) & 5.70 & B\\
CH$_3$OCH$_3$\tablenotemark{i}  & 13$_{5,8}-$13$_{4,9}$ EE & 262.393513 & 82.02 & 148  & 44(3) & 9.2(8) & 5.2(6) & 8.7(2) & 5.12 & B\\
CH$_3$OCH$_3$\tablenotemark{i}  & 13$_{5,8}-$13$_{4,9}$ AA & 262.395111 & 82.02 & 108 & 44(3) & 9.2(8) & 5.2(6) & 10.6(2) & 5.12 & B\\
HNCO & 12$_{0,12}-$11$_{0,11}$ & 263.748625 & 57.19 & 30.0 & 389(6) & 72.8(11) & 5.19(9) & 8.01(4) & 7.65 & B\\
U-line & & 263.7680(3) & & & 72(5) & 12.1(15) & 5.3(7) & & 6.71 &  \\
HC$_3$N\tablenotemark{j} & 29$-$28 & 263.792308 & 132.00 & 404 & 763(6) & 141.4(13) & 5.17(6) & 6.99(2) & 6.71 & B \\
CH$_3$COCH$_3$?\tablenotemark{k} & $22_{6,17}-21_{5,16}$ EA & 263.816703 & 114.36 & 534 & 81(5) & 4.5(3) & 18.4(11) & 6.66(13) & 6.71 & B\\
CH$_2$DOH?\tablenotemark{k} & $2_{2,1}-2_{1,2}$ & 263.818884 & 29.13 & 0.419 & 81(5) & 4.5(3) & 18.4(11) & 9.14(13) & 6.71 & B\\
U-line & & 264.241808(9) &  &  & 129(4) & 24.3(9) & 5.9(3) &  & 6.03 & \\
\enddata  
\tablenotetext{a}{Tentative detections are indicated by a question mark.}
\tablenotetext{b}{Rest frequencies for the identified lines. Fitted rest frequencies assuming the $v_\mathrm{lsr}$ of 8.34 km s$^{-1}$ for the unidentified lines.}
\tablenotetext{c}{B: Broad line, I: Intermediate line, N: Narrow line. See section 3.2.}
\tablenotetext{d}{$I$ and $\Delta v$ are not derived by the Gaussian fit due to the existence of the absorption feature.} 
\tablenotetext{e}{This line may be blended with the $20_{16,4}-19_{16,3}$ E line of HCOOCH$_3$ (245.583970 GHz).}
\tablenotetext{f}{$I$ and $\Delta v$ are not derived by the Gaussian fit because the line is too weak.}
\tablenotetext{g}{$F_{\mathrm{integ}}$, $I$ and $\Delta v$ are not derived by the Gaussian fit because of heavy blending. The K-doublet lines are also unresolved.}
\tablenotetext{h}{This line is blended with the $24_{1,23}-23_{2,22}$ EE  line of CH$_3$COCH$_3$ (248.6815860 GHz).}
\tablenotetext{i}{$F_{\mathrm{integ}}$, $I$ and $\Delta v$ are values for the blended lines.} 
\tablenotetext{j}{This line is likely blended with the $5_{1,5}-4_{2,4}$ A$^{+}$ $v_t = 1$ line of CH$_3$OH (263.7938560 GHz).}
\tablenotetext{k}{These lines are blended with each other. Furthermore, these lines are blended with the $22_{6,17}-21_{5,16}$ AE (263.8167120 GHz), $22_{5,17}-21_{6,16}$ AE (263.8167120 GHz), and $22_{5,17}-21_{6,16}$ EA (263.8167027 GHz) lines of CH$_3$COCH$_3$.}
\end{deluxetable}

\begin{deluxetable}{rccc}
\tablecaption{Column Densities and the Fractional Abundances Relative to H$_2$\tablenotemark{a}}
\tablecolumns{4}
\tablenum{2}
\tablewidth{0pt}
\tablehead{
\colhead{Molecule} & \colhead{Column Density /10$^{14}$ cm$^{-3}$} & \ \ & \colhead{Fractional Abundance /10$^{-10}$}
}
\startdata
HCOOCH$_3$\tablenotemark{b,c} & 26(3) & & 46(5) \\
CH$_3$CHO\tablenotemark{d} & 14(2) & & 24(4)\\
NH$_2$CHO & 2.4(2) &  & 4.3(4) \\
HNCO & 96(10) & & 170(17) \\
HCOOH\tablenotemark{e} & 27(3) & & 47(5)\\
c-C$_3$H$_2$\tablenotemark{f} & $>0.80(8)$ &  & $>1.41(15)$ \\
SO$_2$ & 16.9(17) & & 30(3) \\
SO & 13.6(14) &  &  24(2) \\
CS\tablenotemark{f} & $>5.4(5)$ & & $>9.6(10)$\\
& \vspace{-10pt} & \\
\hline
\multicolumn{4}{c}{Tentative Detection}\\
CH$_3$OCH$_3$ & 19(2) & & 34(4)\\
CH$_3$COCH$_3$\tablenotemark{b} & 4.7(5) & & 8.4(10)\\
C$_2$H$_5$OH & 21(3) & & 38(5) \\
C$_2$H$_5$CN & 0.96(15) & & 1.7(3) \\
\enddata
\tablenotetext{a}{The excitation temperature is assumed to be 100 K.}
\tablenotetext{b}{Derived from the $20_{5,16}-19_{5,15}$ E line (249.031 GHz) and the $20_{5,16}-19_{5,15}$ A line (249.047 GHz).}
\tablenotetext{c}{The vibrationally excited states are not considered in the partition function.}
\tablenotetext{d}{Derived from the $14_{1,14}-13_{1,13}$ E line (260.530 GHz).}
\tablenotetext{e}{Derived from the $12_{0,12}-11_{0,11}$ line (262.103 GHz).}
\tablenotetext{f}{The lower limit is estimated due to the absorption feature.}
\end{deluxetable}

\begin{deluxetable}{lccccccc}
\tablewidth{0pt}
\tablenum{3}
\tablecolumns{8}
\tablecaption{Fractional Abundances of COMs\tablenotemark{a}}
\tablehead{
\colhead{Molecule} & \colhead{B335\tablenotemark{b}} & \colhead{IRAS 16293-2422\tablenotemark{c}} & \colhead{IRAS 4A} & \colhead{IRAS 2A\tablenotemark{d}} 
}
\startdata
$X$(HCOOH) & 4.7 & $\lesssim 0.3$ &  & \\
$X$(CH$_3$CHO) & 2.4 & 3 &  & \\
$X$(HCOOCH$_3$) & 4.6 & 9 & 1.4\tablenotemark{d} & 13\\
$X$(CH$_3$OCH$_3$) & 3.4 & 40 & 0.85\tablenotemark{d} & 8.2\\
$X$(HNCO) & 17 &  & 0.8\tablenotemark{e} & \\
$X$(NH$_2$CHO) & 0.4 & 0.6 & 0.2\tablenotemark{e} & 2.3 \\
$X$(C$_2$H$_5$OH) & 3.8 & $\lesssim$ 5 & 1.2\tablenotemark{d} & 10\\
$X$(C$_2$H$_5$CN) & 0.2 & $\lesssim$ 0.2 & 0.062\tablenotemark{d} & 0.24\\
\enddata
\tablenotetext{a}{$X$ represents the fractional abundance relative to H$_2$ in unit of 10$^{-9}$.}
\tablenotetext{b}{The temperature is assumed to be 100 K.} 
\tablenotetext{c}{\citet{jaber2014} }
\tablenotetext{d}{\citet{taquet2015}}
\tablenotetext{e}{\citet{ana2015}}
\end{deluxetable}


\begin{thebibliography}{}
\bibitem[Adande et al.(2013)]{adande2013} Adande, G. R., Woolf N. J., \& Ziurys, L. M. 2013 Astrobiology, 13,  439
\bibitem[Bacmann et al.(2012)]{bacmann2012} Bacmann, A., Taquet, V., Faure, A., Kahane, C., \& Ceccarelli, C.bacmann 2012, A\&A, 541, L12
\bibitem[Bisschop et al.(2007)]{bisschop2007} Bisschop, S. E., J\o rgensen, J. K., van Dishoeck, E. F., \& de Wachter, E. B. M. 2007, A\&A, 465, 913
\bibitem[Bottinelli et al.(2004)]{bottinelli2004} Bottinelli, S., Ceccarelli, C., Neri, R., et al. 2004, ApJ, 617, L69
\bibitem[Cazaux et al.(2003)]{cazaux2003} Cazaux, S., Tielens, A. G. G. M., Ceccarelli, C., et al. 2003, ApJ, 593, L41  
\bibitem[Chandler et al.(1993)]{chandler1993}Chandler, C. J. \& Sargent, A. I. 1993, ApJ, 414, L29
\bibitem[Codella et al.(2016)]{codella2016} Codella, C., Ceccarelli, C., Cabrit, S., et al.\ 2016, \aap, 586, L3
\bibitem[Evans et al.(2005)]{evans2005} Evans, N.~J., II, Lee, J.-E., Rawlings, J.~M.~C., \& Choi, M.\ 2005, \apj, 626, 919
\bibitem[Evans et al.(2015)]{evans2015} Evans, N. J., II, Di Francesco, J., Lee, J., et al.  2015, ApJ, 814, 22
\bibitem[Friedel et al.(2005)]{friedel2005} Friedel, D. N., Snyder, L. E., Remijan, A. J., \& Turner, B.E. 2005, ApJ, 632, L95
\bibitem[Friedel \& Snyder (2008)]{friedelandsnyder2008} Friedel, D. N. \& Snyder, L. E. 2008, ApJ, 672, 962
\bibitem[Harvey et al.(2001)]{harvey2001} Harvey, D. W. A., Wilner, D. J., Lada, C. J., et al. 2001, ApJ, 563, 903
\bibitem[Hirano et al.(1988)]{hirano1988} Hirano, N., Kameya, O., Nakayama, M., \& Takakubo, K. 1988, ApJ, 327, L69
\bibitem[Hirano et al.(1992)]{hirano1992} Hirano, N., Kameya, O., Kasuga, T., \& Umemoto, T. 1992, ApJ, 390, L85
\bibitem[Jaber et al.(2014)]{jaber2014} Jaber, A. A., Ceccarelli, C., Kahane, C., \& Caux, E. 2014, ApJ, 791, 29
\bibitem[J{\o}rgensen et al.(2005)]{jorgensen2005} J{\o}rgensen, J.~K., Bourke, T.~L., Myers, P.~C., et al.\ 2005, \apj, 632, 973 
\bibitem[Kahane et al.(2013)]{kahane2013} Kahane, C., Ceccarelli, C., Faure, A., \& Caux, E. 2013, ApJ, 763, L38
\bibitem[Kauffmann et al.(2008)]{kauffmann2008} Kauffmann, J., Bertoldi, F., Bourke, T.~L., Evans, N.~J., II, \& Lee, C.~W.\ 2008, \aap, 487, 993
\bibitem[Keene et al.(1980)]{keene1980} Keene, J., Hildebrand, R.H., Whitcomb, S. E., \& Harper, D. A. 1980, ApJ, 240, L43
\bibitem[Kuan et al.(2004)]{kuan2004} Kuan, Y., Huang, H., Charnley, S. B., et al. 2004, ApJ, 616, L27
\bibitem[Lindberg et al.(2015)]{lindberg2015} Lindberg, J. E., J\o rgensen, J. K., Watanabe, Y., et al. 2015, A\&A, 584, A28
\bibitem[L$\acute{\mathrm{o}}$pez-Sepulcre et al.(2015)]{ana2015} L$\acute{\mathrm{o}}$pez-Sepulcre, A., Jaber, A. A., Mendoza, E., et al. 2015, MNRAS, 449, 2438
\bibitem[M$\ddot{\mathrm{u}}$ller et al.(2005)]{muller2005}M$\ddot{\mathrm{u}}$ller, H. S. P., Schl$\ddot{\mathrm{o}}$der, F., Stutzki, J., \& Winnewisser, G. 2005, JMoSt, 742, 215
\bibitem[$\ddot{\mathrm{O}}$berg et al.(2011)] {oberg2011} $\ddot{\mathrm{O}}$berg, K., van der Marel, N., Kristensen, L. E., \& van Dishoeck, E. F. 2011, ApJ, 740, 14
\bibitem[Olofsson \& Olofsson (2009)]{olofsson2009} Olofsson, S. \& Olofsson, G. 2009, A\&A, 498, 455
\bibitem[Oya et al.(2016)]{oya2016} Oya, Y., Sakai, N., L$\acute{\mathrm{o}}$pez-Sepulcre, A., et al. 2016, ApJ, 824, 88
\bibitem[Peng et al.(2013)]{peng2013} Peng, T.-C., Despois, D., Brouillet, N., et al. 2013, A\&A, 554, A78
\bibitem[Pickett et al.(1998)]{pickett1998} Pickett, H. M., Poynter, R. L., Cohen, E. A., et al. 1998, JQSRT, 60, 883
\bibitem[Richard et al.(2013)]{richard2013} Richard, C., Margul$\grave{\mathrm{e}}$s, L., Caux, E., et al. 2013, A\&A, 552, A117
\bibitem[Sakai et al.(2006)]{sakai2006} Sakai, N., Sakai, T., \& Yamamoto, S.\ 2006, \pasj, 58, L15
\bibitem[Sakai et al.(2008)]{sakai2008} Sakai, N., Sakai, T., Hirota, T., \& Yamamoto, S. 2008, ApJ, 672, 371
\bibitem[Sakai et al.(2009)]{sakai2009} Sakai, N., Sakai, T., Hirota, T., Burton, M., \& Yamamoto, S. 2009, ApJ, 697, 769
\bibitem[Sakai et al.(2010)]{sakai2010} Sakai, N., Sakai, T., Hirota, T., \& Yamamoto, S. 2010, ApJ, 722, 1633
\bibitem[Sakai \& Yamamoto (2013)]{sakaiandyamamoto2013} Sakai, N. \& Yamamoto, S. 2013, Chem. Rev., 113, 8981
\bibitem[Saladino et al.(2012)]{saladino2012} Saladino, R., Botta, G., Pino, S., Costanzo, G., \& Di Mauro, E. 2012, Chem. Soc. Rev., 41, 5526  
\bibitem[Shirley et al.(2011)]{shirley2011} Shirley, Y.~L., Huard, T.~L., Pontoppidan, K.~M., et al.\ 2011, \apj, 728, 143
\bibitem[Spezzano et al.(2016)]{spezzano2016} Spezzano, S., Bizzocchi, L., Caselli, P., Harju, J., \& Br$\ddot{\mathrm{u}}$nken, S. 2016, A\&A, 592, L11
\bibitem[Sugimura et al.(2011)]{sugimura2011} Sugimura, M., Yamaguchi, T., Sakai, T., et al. 2011, PASJ, 63, 459
\bibitem[Taquet et al.(2015)]{taquet2015} Taquet V., L$\acute{\mathrm{o}}$pez-Sepulcre, A., Ceccarelli, C., et al. 2015, ApJ, 804, 81
\bibitem[Ward-Thompson et al.(2000)]{ward2000} Ward-Thompson, D., Zylka, R., Mezger, P. G., \& Sievers, A. W. 2000, A\&A, 335, 1122
\bibitem[Watanabe et al.(2012)]{watanabe2012} Watanbe, Y., Sakai, N., Lindberg, J. E., et al. 2012, ApJ, 745, 126
\bibitem[Wilner et al.(2000)]{wilner2000} Wilner, D. J., Myers, P. C., Mardones, D., \& Tafalla, M. 2000, ApJ, 544, L69
\bibitem[Yen et al.(2015)]{yen2015} Yen, H., Takakuwa, S., Koch, P. M., et al. 2015, ApJ, 812, 129
\bibitem[Zhou et al.(1993)]{zhou1993} Zhou, S., Evans, N. J., II, Koempe, C., \& Walmsley, C. M. 1993, ApJ, 404, 232
\end{thebibliography}
\end{document}